\documentclass[twocolumn,pre,showpacs,preprintnumbers,amsmath,amssymb,superscriptaddress]{revtex4}
\usepackage{graphicx}
\usepackage{bm}
\usepackage{longtable}

\begin{document}

\markboth{Burioni et al.}{Philosophical Magazine}


\title{Local and average behavior in inhomogeneous superdiffusive media}

\author{ Alessandro Vezzani$^{\rm a,b}$$^{\ast}$, Raffaella Burioni$^{\rm b,c}$, Luca Caniparoli$^{\rm d}$ and Stefano Lepri$^{\rm e}$  \thanks{$^\ast$Corresponding author. Email: vezzani@fis.unipr.it \vspace{6pt}}
\\\vspace{6pt}  $^{\rm a}${{Centro S3, CNR-Istituto di Nanoscienze, Via Campi 213A, 41125 Modena Italy} $^{\rm b}${{Dipartimento di Fisica, Universit\`a degli Studi di Parma, viale G.P.Usberti 7/A, 43100 Parma, Italy}}; $^{\rm c}${{INFN, Gruppo Collegato di Parma, viale G.P. Usberti 7/A, 43100 Parma, Italy}}; $^{\rm d}${{International School for Advanced Studies SISSA, via Beirut 2/4, 34151, Trieste, Italy}}; $^{\rm e}${{Istituto dei Sistemi Complessi, Consiglio Nazionale delle Ricerche, via Madonna del Piano 10, I-50019 Sesto Fiorentino, Italy}}}\\\vspace{6pt}\received{\today} }

\begin{abstract} 

We consider a  random walk on one-dimensional inhomogeneous graphs built from Cantor fractals.
Our study is motivated by recent experiments that demonstrated superdiffusion of light in
complex disordered materials, thereby termed  L\'evy glasses. We introduce a geometric
parameter $\alpha$ which plays a role analogous to the exponent characterizing the step length
distribution in random systems. We study the large-time behavior of both local and average
observables; for the latter case, we distinguish two different types of averages, respectively
over the set of all initial sites and over the scattering sites only. The ``single long jump
approximation" is applied  to  analytically determine the different asymptotic
behaviours as a function of $\alpha$  and to understand their origin.  We also discuss the
possibility  that the root of the mean square displacement and the characteristic length of the
walker distribution may grow according to different power laws; this anomalous behaviour is
typical of processes characterized by L\'evy statistics and here, in particular, it is shown to
influence average quantities.

\end{abstract}

\maketitle
\section{Introduction}
The laws of Brownian motion crucially relies on the hypothesis  that the
steps for the diffusing particle are  small (with finite
variance) and uncorrelated.  Whenever these assumptions are violated, 
the standard diffusion picture  breaks down and anomalous phenomena emerge
\cite{bouchaud,klages}. 

In particular, transport processes where the step length distribution has a
diverging variance have been theoretically studied in detail. Among those, one of the most interesting 
is the so-called L\'evy walks  \cite{Blumen1989,Klafter1990}, in which particles
perform independent steps $l$ at constant velocity, with a  distribution $\lambda(l)$
following an algebraic tail $\sim l^{-(1+\alpha)}$. Such a distribution is said to be
heavy-tailed and it  is known to have a diverging variance for  $\alpha< 2$. Since
transport is thereby dominated by very long steps, the mean square displacement
increases faster than linearly  with time, hence the name superdiffusion. 

Among the many possible experimental applications, our work is motivated 
by the recent realization of materials termed \emph{L\'evy glasses}, where
light rays propagate through an assembly of transparent spheres embedded
in a scattering medium \cite{Barthelemy08,Bertolotti10}. If the diameter
of the spheres is power-law distributed, light can indeed perform anomalous
diffusion. Owing to their ease of fabrication and tunability, such a 
novel material offers an unprecedented opportunity to study anomalous 
transport processes in a systematic and controllable way.

An important feature of the experimental samples is that the walk is 
correlated: light that has just crossed a large glass microsphere  has a higher
probability of being backscattered at the following step  and thus to perform a
jump of roughly the same length. While the case of uncorrelated jumps is well
understood \cite{ann}, the correlation effects, that are expected to deeply
influence  the diffusion properties \cite{Fogedby94}, are still to be 
characterized. To this aim, quenched L\'evy processes have been studied on one
dimensional systems \cite{klafter,beenakker}. More recently, different aspects
regarding the scaling properties of random-walk distributions, the relations
between  the dynamical exponents and the different average procedures have been
discussed in a common framework \cite{noi2}. 

In order to get a deeper insight on the effect of step-length correlations, a
class of deterministic, one-dimensional models called \textit{Cantor 
graphs} has been introduced \cite{noi}. Random walks on these structures perform 
correlated long jumps induced by the underlying fractal topology. 
As the latter is generated by deterministic rules, diffusion properties
can be studied in a simpler way than in the random case.
Here we extended to this deterministic topologies some of the results proved 
in \cite{noi2} for random structures. In particular, we introduce a geometric parameter 
$\alpha$ which plays the same role as the exponent characterizing  
the step length distribution $\lambda(l)$ in random systems.  Three kinds
of statistical averaging are introduced: (i) a local one, namely the 
average of all trajectories starting from a given initial site
(whose asymptotic behaviour being expected to be independent of the site choice); 
(ii) an averaging over \textit{all} possible initial sites 
of the graph and (iii) averages where \textit{only} scattering sites are considered
as initial conditions. 
In the random case, the differences between such averaging procedures have been 
evidenced in \cite{klafter,noi2}. 
On determinisitc structures, average procedures (i) and (ii)
have been discussed in \cite{noi}. Here we complete the picture studying 
the effect of averaging over scattering sites. We evidence that 
the behavior of the mean square displacement is similar to random case; 
while the probability density displays a more complex structure being 
given by a non-trivial time-dependent superposition
of step functions. We remark that,
in experiments \cite{Barthelemy08,Bertolotti10}, light 
enters the sample with a scattering event and averages of type (iii)
are the most physically sensible quantities to compare with. 
Interestingly, in  L\'evy processes the root of the mean square 
displacement and the characteristic length of the distributions may grow
according to different power laws; here, in particular, this strongly 
anomalous behavior characterizes average quantities.

The paper is organized as follows. In the next section we introduce the directed  Cantor graphs
and we define a simple random walk on these structures.  We then discuss the relevant physical
quantites  and the average procedures. Section 4 is devoted to discuss the scaling hypothesis
and the single long jump approximation, which allows to evaluate the tails of the density
distribution. Finally, in Section 5 we discuss our results evidencing differences among
averaging procedures. In particular for the case (i) and (ii) we review the results presented in 
\cite{noi} presenting new numerical data,
while for the case of average over scattering sites, which has
not been discussed so far for the deterministic structures, we provide a sketch of the
derivation of the dynamical exponents within the single long jump approximation and we 
compare the result with numerical simulations. In general the new simulations evidence 
that the probability density presents different asymptotic behaviors depending on the average 
procedures, clearly supporting both the scaling hypothesis and the single long jump approach. Moreover the asymptotic behavior of the mean square displacement have been 
tested for a wider range of $\alpha$'s.

\section{Random walks on Cantor graphs}

In paper \cite{noi} we have introduced a class of graphs, denoted as directed
Cantor graphs, as a one-dimensional, deterministic counterpart of the
geometric structure of the L\'evy glass materials mentioned above
\cite{Barthelemy08}. Indeed, random walks on the Cantor graphs
display a superdiffusive L\'evy like motion analogous to the the 
one of light in such inhomogeneous glassy material. 

\begin{figure}
\begin{center}
\includegraphics[width=0.5\textwidth]{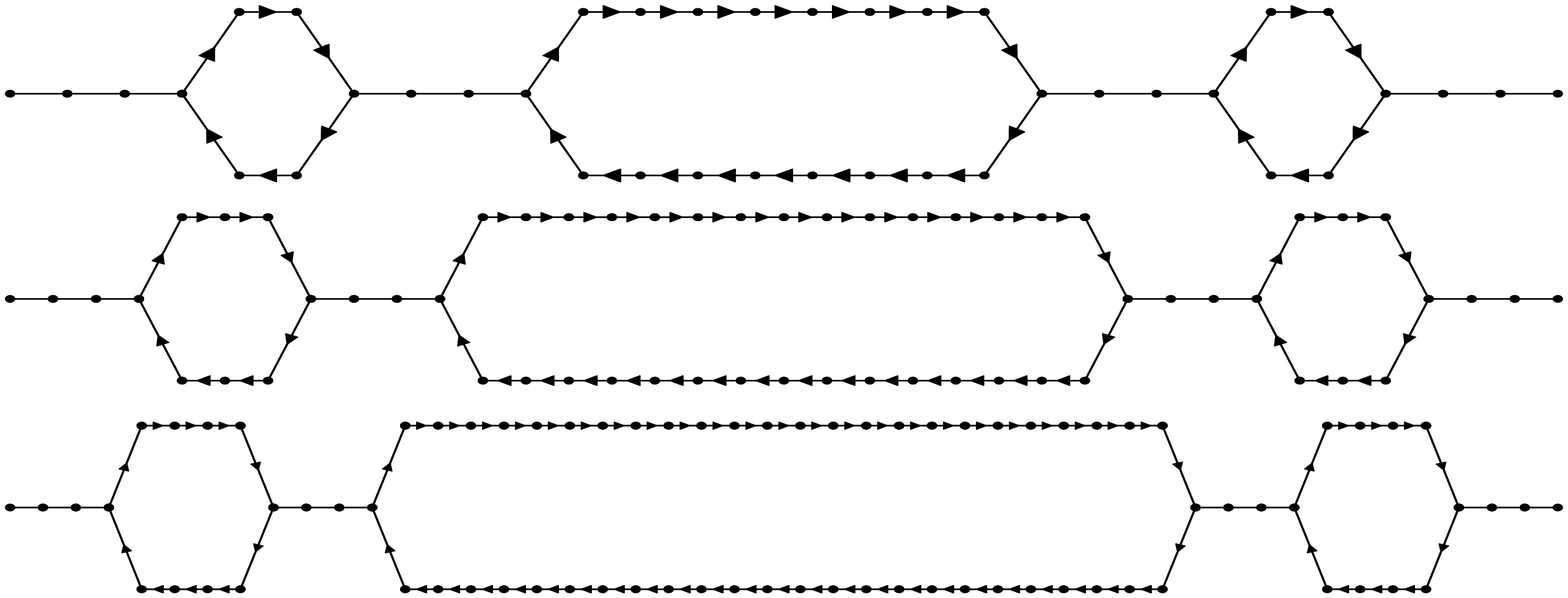}
\caption{The generation $G=3$ of the  graph  with $n_u=3,4,5$, in the 
$n_r=2$ case.}
\label{cantor_walk_var}
\end{center}
\end{figure}

\begin{figure}
\begin{center}
\includegraphics[width=0.5\textwidth]{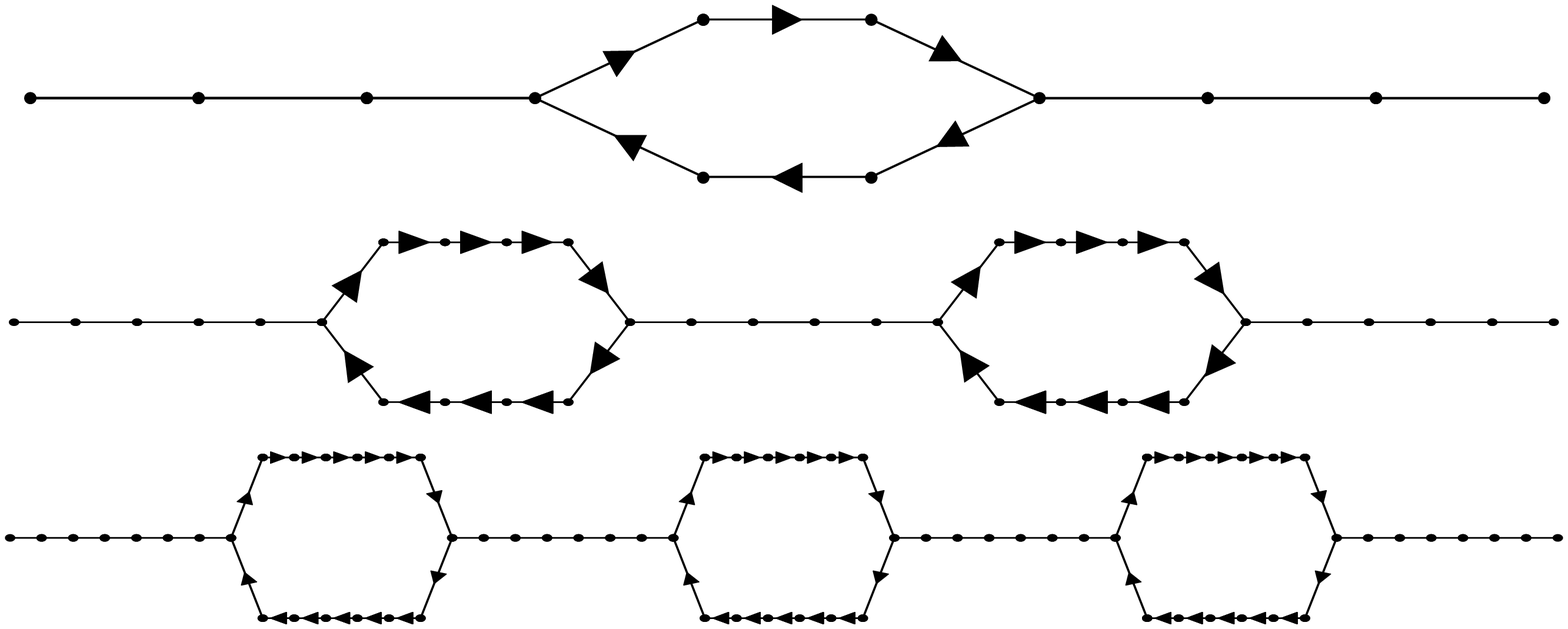}
\caption{Three examples of Cantor graphs of generation $G=2$ with 
$n_r=2,3,4$ and $n_u=3,4,5$.}
\label{cantor_walk_nat}
\end{center}
\end{figure}

The class of graphs we will consider is defined by two parameters, denoted as
$n_r$ and $n_u$, and describing the growth of the fractal from generation $G-1$
to generation $G$. In particular the fractal $\cal G$ of generation $G$ is built
connecting $n_r$ fractals  of generation $G-1$ by $n_r-1$ unidirectional bubbles
of length  $L_G=n_u^{G-1}$, as shown in  Figures 
\ref{cantor_walk_var} and \ref{cantor_walk_nat}.

A simple random walk \cite{rw} is naturally defined on these structures: undirected links  
connect sites in both directions while directed links 
have to be crossed only in the prescribed way.
A site is called \textit{bidirectional} if the walker placed on that site
can move  in both directions (and the two possible moves are performed with 
probability $1/2$).
A site is instead called \textit{unidirectional} if the walker is allowed to move 
only in one direction (in this case the move is performed with probability one, i.e. 
ballistic motion). All the links have unitary length and
are crossed at constant velocity $v$. The number of
bidirectional sites $N_b$ and unidirectional sites $N_u$ present at generation
$G$ is given by:
\begin{equation}
N_{b}=2n_r^G,\qquad 
N_{u} =  
\frac{(n_u-1)n_r^G-\ (n_r-1)n_u^G}{n_r-n_u}+1
\label{nu}
\end{equation}
where we use the convention of counting only once the couple of sites of the
bubbles at the same distance from the origin. The total number of sites at
generation $G$ is hence $N_G=N_b+N_u$. In the following, an important parameter
for the description of the structure will be the ratio
$\alpha=\log(n_r)/\log(n_u)$. For $\alpha<1$, the graphs are called {\itshape
slim}, as the fraction of bidirectional sites vanishes in the thermodinamic 
limit, i.e. $\lim_{G \to \infty}N_b/N_G=0$. 
On the contrary, for  $\alpha>1$ graphs are called
{\itshape fat}, since a finite fraction of bidirectional sites is present, and
$\lim_{G \to \infty}N_b/N_G>0$.

\section{Physical quantities and averages}

Let $P_i(r,t)$ be the probability of arriving at distance $r$ starting from $i$ in $t$ steps. In general, $P_i(r,t)$ depends on the starting site $i$. However for large enough space
scales (i.e. $r$ much larger than the distance between $i$ and an origin $i=1$)  the asymptotic
properties of $P_i(r,t)$ are expected to be site independent and hence to describe a property
of the whole graph, i.e. $P_i(r,t)\sim  P_1(r,t)$. In particular, the asymptotic behaviour of
the mean square displacement
\begin{equation}
\langle r_i^2(t)\rangle = \int_{0}^{vt} P_i(r,t) r^2 dr
\label{r2loc}
\end{equation}
should be independent of $i$.
The integration cutoff in (\ref{r2loc}) is provided by the fact that the walker
covers at most a distance $vt$ in a time $t$ ($v=1$ in the following). 
Another important quantity whose
asymptotic behaviour depends only on the large scale topology of the
structure is the resistivity $R_i(r)$ i.e the number of bidirectional sites
whose distance from $i$ is smaller than $r$. 

On inhomogeneous structures, average and local properties are in general
different \cite{burioni}, and in structures characterized by long tails,
different averaging procedures can yield different results as well
\cite{klafter,noi2}. In particular, on Cantor graphs one can distinguish
between the average over all starting sites, i.e.:
\begin{equation}
\bar P(r,t)= \lim_{G \to \infty}\frac{ \sum_{i\in{\cal G}} P_i(r,t)}{N_G} 
\label{avgtot}
\end{equation}
and the averages on  processes beginning with a scattering event:
\begin{equation}
\tilde P(r,t)= \lim_{G \to \infty}\frac{ \sum_{i\in{\cal G}_b} P_i(r,t)}{N_b} 
\label{avgsct}
\end{equation}
where ${\cal G}_b$ is the set of the bidirectional sites belonging to graphs of
generation $G$. The same averaging procedures can be introduced also for
different quantities. One can consider the average resistivity $\bar{R}(r)$ and
$\tilde{R}(r)$ and, as in (\ref{r2loc}), one can also define the average mean
square distances $\langle \bar{r}^2(t)\rangle$ and $\langle
\tilde{r}^2(t)\rangle$. 

Since the resistivity can be evaluated by simply counting the number of sites in a given
generation, one obtain the following asymptotic behaviours:
\begin{equation}
R_1(r)\sim \tilde{R}(r)
\sim\left\{\begin{array}{lc} r^\alpha, & \mathrm{if}\ \alpha<1 \\
\\
r & \mathrm{if}\ \alpha \geq  1
\end{array}\right.
\label{Rloc}
\end{equation}

\begin{equation}
\bar{R}(r)\sim
\left\{\begin{array}{lc} 0 & \mathrm{if}\ \alpha<1 \\
\\
r & \mathrm{if}\ \alpha \geq  1.
\end{array}\right.
\label{Rav}
\end{equation}
These results represent the deterministic analog of the expressions for the
resistivity obtained in a random sample in  \cite{beenakker}.

\section{The scaling hypothesis and the single long jump approximation}

The most general scaling hypothesis for the probabilities $P_1(r,t)$ is:
\begin{equation}
P_1(r,t)=\ell_1^{-1}(t)f_1(r/\ell_1(t))+g_1(r,t) 
\label{sal}
\end{equation}
with  a convergence in probability
\begin{equation}
\lim_{t\to\infty}\int_0^{t} |P_1(r,t)-\ell^{-1}_1(t)f_1(r/\ell_1(t))|dr=0
\label{sal2}
\end{equation}
The leading contribution to $P_1(r,t)$  is hence  
$\ell_1^{-1}(t)f_1(r/\ell_1(t))$ which is significantly different from 
zero only for $r \lesssim \ell_1(t)$. 
The subleading term $g_1(r,t)$, with $\lim_{t \to \infty}\int |g_1(r,t)| dr=0$ describes the behavior 
at larger distances, i.e. $\ell_1(t)\ll r < t$. Notice that, 
if $g_1(r,t)$ does not vanish rapidly enough, 
it can nevertheless provide important contributions to 
$\langle r_1^2(t)\rangle$.
The same scaling ansatz should be valid also for the average 
probabilities $\bar{P}(r,t)$ and $\tilde{P}(r,t)$ by introducing suitable averaged 
scaling length and  scaling functions $\bar{f}(r/\bar{\ell})$, $\tilde{f}(r/\tilde{\ell})$, and suitable averaged long distance corrections $\bar{g}(r,t)$, $\tilde{g}(r,t)$.

In \cite{noi,cates} it has been proved that the growth of the characteristic
length can be directly related to the growth of the resistance. 
We define the exponent describing the growth of the correlation length as:
\begin{equation}
\ell_1(t)\sim t^{d_s/2}.
\label{sal3}
\end{equation}
so that in analogy with standard definition of random walks \cite{Orbach}, we get
$P_1(0,t)\sim t^{-d_s/2}$.
Then using the scaling relations proved in \cite{noi,cates} we obtain
\begin{equation}
R_1(r)\sim r^{2/d_s-1}.
\label{scalR}
\end{equation}
Analogous relations hold for the average quantities, in terms of the average scaling lengths. Introducing the known results for the resistivity (\ref{Rloc},\ref{Rav}), one obtains the following behaviors for the scaling lenghts:
\begin{equation}
\ell_1(t)\sim \tilde{\ell}(t)
\sim \left\{\begin{array}{lc} t^{\frac 1{1+\alpha}}, & \mathrm{if}\ \alpha<1 \\
\\
t^{\frac 1 2} & \mathrm{if}\ \alpha \geq  1
\end{array}\right.
\label{elloc}
\end{equation}

\begin{equation}
\bar{\ell}(t)\sim
\left\{\begin{array}{lc} t & \mathrm{if}\ \alpha<1 \\
\\
t^{\frac 1 2} & \mathrm{if}\ \alpha \geq  1
\end{array}\right.
\label{elav}
\end{equation}

Let us now discuss the behaviour of the mean square displacements. When only
lengths of order $ r \lesssim \ell(t)$ provide significant contributions to
the integral (\ref{r2loc}), the standard relation $\langle r^2 (t) \rangle \sim
\ell^2(t)$ holds and the asymptotic behaviour coincides with those given by 
(\ref{elloc},\ref{elav}). However it is known that, in
presence of long tailed distributions, anomalies with respect to this behaviour
can be present. In  particular it has been evidenced in \cite{klafter,noi2} that
a key role is played by long jumps, leading the walker to a distance $r\gg
\ell(t)$. In the random case, consideration of
a single long jump actually accounts for the asymptotic behaviour \cite{noi2}. 
Indeed, these processes can give rise to two different types of corrections to
$P(r,t)$.  First, they can  produce a zero-measure function $g(r,t)$,  providing
a significant contribution to  $\langle r^2 (t) \rangle$; second, the scaling
function $f(x)$ can feature a long tail, breaking the proportionality between
$\langle r^2 (t) \rangle$ and $ \ell^2(t)$. Here we will evidence that these
anomalies, originating from the single long jump, are also present in the
deterministic graphs and they are deeply influenced by the averaging procedures.

\section{Results}

\subsection{Local behaviour}

\begin{figure}
\begin{center}
\includegraphics[width=0.5\textwidth,clip]{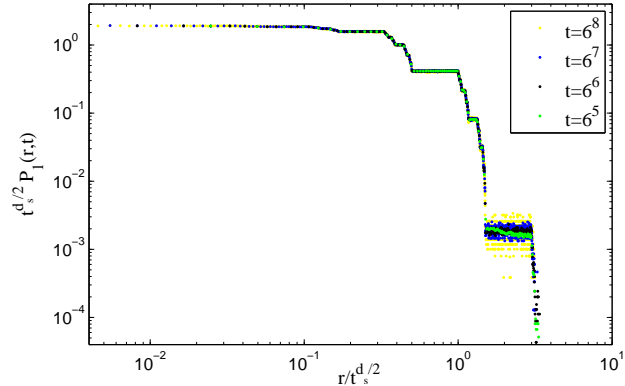}
\caption{Dynamical scaling of the probability for initial  site 
$i=1$ on the graph obtained with $n_r=2$ and $n_u=3$. Note
the fast decay of the scaling function.  The stepwise structure 
of the scaling function is due to the fractality of the graph.}
\label{dynscal}
\end{center}
\end{figure}

\begin{figure}
\begin{center}
\includegraphics[width=0.5\textwidth,clip]{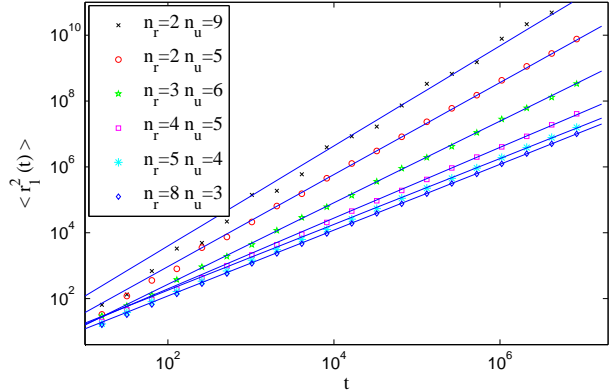}
\caption{Growth of the mean square displacements for initial 
site $i=1$. Power-law fits give exponents in very good agreement 
with the theoretical values, in equation (\ref{msqloc}). } 
\label{x2}
\end{center}
\end{figure}

Let us first consider the local properties. In this situation we can focus on 
processes starting from the origin $i=1$ of the graph. Indeed,  if $r$ is much
larger than the distance between $i$ and $1$, we expect that $P_i(r,t)$ should
behave as $P_1(r,t)$ and hence the asymptotic behaviour should be the same for
any starting point. When starting from the origin, in a time $t$ the walker
typically covers a distance $\ell_1(t)$ and, in the deterministic graphs,
within such distance there are no bubbles of length larger than  $\ell_1(t)$. 
Therefore, long jumps do not occur and 
$\langle r^2_1(t)\rangle\sim  \ell_1^2(t)$. Therefore \cite{noi}
\begin{equation}
\langle r^2_1(t)\rangle
\sim \left\{\begin{array}{lc} t^{\frac 2{1+\alpha}}, & \mathrm{if}\ \alpha<1 \\
\\
t & \mathrm{if}\ \alpha \geq  1
\end{array}\right.
\label{msqloc}
\end{equation}
Figure \ref{dynscal} reports, an example the probability density $P_1(r,t)$  
obtained by a Montecarlo simulation. The data provide a clear evidence  that the scaling 
function presents a fast decay confirming our hypothesis of no long jump. 
As explained in \cite{noi}  the fractal structure give rise to log-periodic oscillations 
which can be discarded considering peculiar sequence of times (in this case $t=6^k$) ,
However, such oscillations do not change the general framework of the scaling
hypothesis.  The growth of $\langle r_1^2(t) \rangle$ is plotted in Figure \ref{x2}, 
the continuous lines represent the expected behaviors  (\ref{msqloc}).

\subsection{Averages over all sites}

\begin{figure}
\begin{center}
\includegraphics[width=0.5\textwidth,clip]{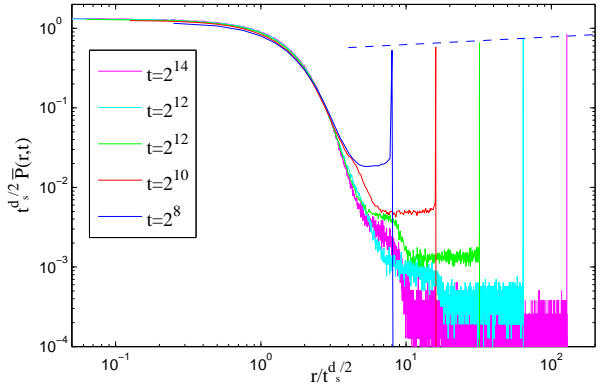}
\caption{Dynamical scaling of $\bar{P}(r,t)$ in the average case. The data refer to the case $n_r=7$ $n_u=4$. Ballistic peaks at $r= t$ scale as $t^{1-\alpha}$  (dashed line), while for $r \lesssim \bar{\ell}(t)$ the scaling hypothesis is well satisfied.} 
\label{pscaling_73_medio}
\end{center}
\end{figure}

\begin{figure}
\begin{center}
\includegraphics[width=0.5\textwidth,clip]{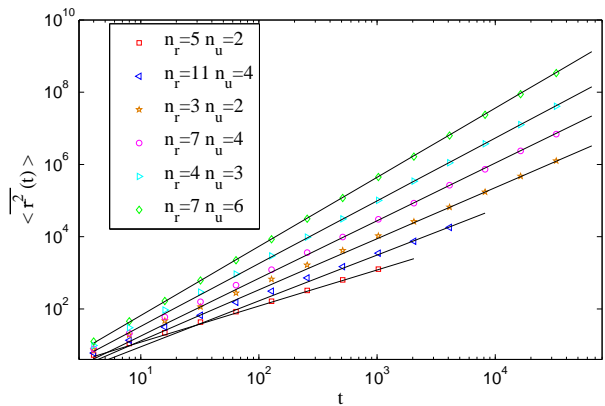}
\caption{Growth of the mean square displacements when average 
over scattering over the whole sample is considered. We compare 
the numerical results with the theoretical preditions (\ref{r2avg}) in both the regimes  $1<\alpha<2$ and $\alpha>2$.} 
\label{x2_73_92}
\end{center}
\end{figure}

When averaging over all the sites, it has been evidenced \cite{noi}  that 
for $\alpha<1$ the motion is always ballistic, while 
for $\alpha>1$ the situation is much more complex, since the walker can perfom 
a single jump much larger than $\bar{\ell}(t)$. Tipically, such a long jump 
occurs at the first step because, with a random choice of the starting point, 
the probability of belonging to a large bubble is much larger  at $t=0$ than 
during the rest of the evolution. In this situation, one can estimate the zero 
measure correction to $\bar{P}(r,t)$ obtaining 
$\bar{g}(r,t)\sim t^{-\alpha+1} \delta(r-t)$, i.e. a peak 
associated with the ballistic motion of the particle, weighted by a factor 
$ t^{-\alpha+1} $ representing the probability of belonging to a bubble larger than $t$
at the initial time.  The behavior of $P(r,t)$ is illustrated in the simulations of figure 
\ref{pscaling_73_medio}  evidencing the presence of a scaling regime for 
$r \lesssim \bar{\ell}(t)$ and of balistic peaks at large $r$ whose height evolves as 
$t^{-\alpha+1}$ (dased line).
Even if $\bar{g}(r,t)$ provides a 
subleading contribution to $\bar{P}(r,t)$, the integral in  (\ref{r2loc}) 
is dominated by $\bar{g}(r,t)$ for $1<\alpha<2$ and by 
$\bar{f}(r/\bar{\ell}(t))$ only when $\alpha>2$. 
Therefore the behaviour of the mean square displacement is
\begin{equation}
\langle \bar{r}^2(t)\rangle
\sim \left\{\begin{array}{lc} t^{3-\alpha}, & \mathrm{if}\ 1<\alpha<2 \\
\\
t & \mathrm{if}\ \alpha \geq  2
\end{array}\right.
\label{r2avg}
\end{equation}
while the motion is purely ballistic for $\alpha<1$. Figure \ref{x2_73_92} 
shows that the predicted exponents (\ref{r2avg}) are well verified.

\subsection{Averages over bidirectional sites}

In \cite{noi2} it has been shown that, for random systems, averages over 
bidirectional scattering sites can provide different results with respect 
to averages over all starting points and a scaling approach has been 
discussed, based on the single long jump approximation. Here we introduce 
an analogous argument for the deterministic graph. 
\begin{figure}
\begin{center}
\includegraphics[width=0.5\textwidth,clip]{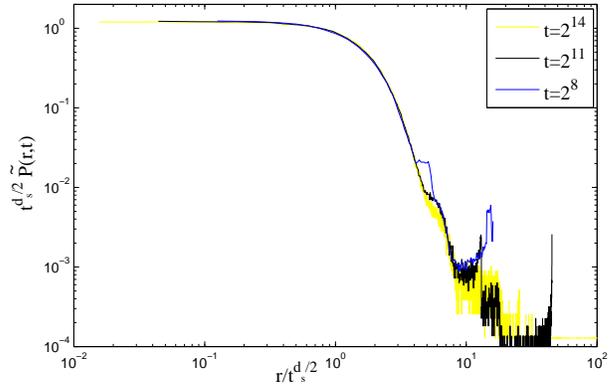}
\caption{The dynamical scaling of $\tilde{P}(r,t)$ in the case of averages 
over bidirectional sites. Here we plot the case $n_r=4$ and $n_u=3$. For $r \lesssim \tilde{\ell}(t)$ the scaling hypothesis (\ref{sal}) is very well verified with $\tilde{\ell}(t)$ growing as indicate in (\ref{elloc}). For larger $r$ the behaviour is much more complicated since the tails  are composed by a superpositions of step functions.}
\label{scalPmeansc}
\end{center}
\end{figure}

\begin{figure}
\begin{center}
\includegraphics[width=0.5\textwidth,clip]{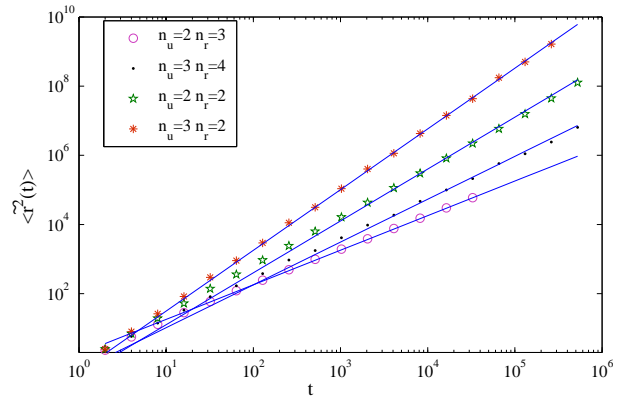}
\caption{Growth of the mean square displacement when averaged 
over bidirectional sites is consider. The results are compared with the 
theoretical predictions (\ref{msqav}) given by the continuous lines. 
Different values of $n_r$ and $n_u$ are considered, showing that the predictions  (\ref{msqav}) are verified in the different regimes of $\alpha$'s.} 
\label{r2meansc}
\end{center}
\end{figure}

In the case of average over bidirectional sites, the single long jump does 
not occur necessarily at the first step. In particular for the deterministic 
graph the probability of performing a jump of length 
$L_k=n_u^k\gg \tilde{\ell}(t)$ in a time $t$ is $N(t) n_r^{-k}$ where $N(t)$ 
is the number of bidirectional sites visited by the walker in a time $t$ and 
$n_r^{-k}$ is the probability that a bidirectional site belongs to a bubble 
of length $L_k$. Discarding this long jump, the distance crossed by the 
walker in a time $t$ is of order $\tilde{\ell}(t)$, and therefore,  according 
to the behaviour of the resistivity described by equations \ref{Rloc},  
$N(t)\sim\tilde{\ell}(t)^\alpha$ for $\alpha<1$ and 
$N(t)\sim\tilde{\ell}(t)$ for $\alpha \geq 1$. The main difference with 
respect to the random case is that now the only possible lengths of the long 
jumps  are given by the sizes $L_k$ of the bubbles in the fractal. 
Hence, for $\tilde{\ell}(t)<r<t$, $\tilde{P}(r,t)$ is  a complex step 
function where both integers $n_u$ and $n_r$ plays a non trivial role. 
However, the contribution to the mean square displacement can be evaluated as 
follows:
\begin{equation}
N(t)\left(\sum_{\tilde{\ell}(t)<L_k<t} \frac{n_u^{2k}}{n_r^{k}}+t^2 \sum_{L_k>t} \frac{1}{n_r^{k}}\right)
\label{msqav1}
\end{equation}
where the first sum is related to the bubbles of length $L_k$ ($\tilde{\ell}(t)<L_k<t$) providing a contribution to the mean square displacements of order $L_k^2=n_u^{2k}$, while the second sum comes from the bubbles of length larger than $t$, providing a contribution $t^2$. Expanding equation \ref{msqav1} for large times one obtains the asymptotic behaviours 
\begin{equation}
N(t)\left(\sum_{\tilde{\ell}(t)<L_k<t} \frac{n_u^{2k}}{n_r^k}+ t^2 \sum_{L_k>t} \frac{1}{n_r^k} \right)\sim t^{\frac{2+2 \alpha+\alpha^2}{1+\alpha}}
\label{msqav2}
\end{equation}
for $\alpha<1$ and
\begin{equation}
N(t)\left(\sum_{\tilde{\ell}(t)<L_k<t} \frac{n_u^{2k}}{n_r^k}+t^2 \sum_{L_k>t} \frac{1}{n_r^k} \right)\sim t^{5/2-\alpha}
\label{msqav3}
\end{equation}
for $\alpha>1$. The first expression is always dominant with respect $\bar{\ell}(t)$, while the second expression becomes subleading for $\alpha> 3/2$. The overall behaviour of the mean square displacement is summarized as  follows:
\begin{equation}
\langle \tilde{r}^2 (t) \rangle \sim
\begin{cases}
t^{\frac{2+2 \alpha-\alpha^2}{1+\alpha}} & \mathrm{if}\ 0<\alpha<1 \\
t^{\frac{5}{2}- \alpha} & \mathrm{if}\ 1 \leq \alpha \leq 3/2 \\
t & \mathrm{if}\ 3/2 < \alpha  
\label{msqav}
\end{cases}
\end{equation}

Equations (\ref{msqav}) extend the results of the random case to the 
deterministic topology described by the directed Cantor graphs. Clearly the 
complex shape of the scaling function (\ref{msqav1}) determines the presence 
of logperiodic oscillations superimposed to (\ref{msqav}), which is a a typical behaviour 
of fractal structures \cite{periodic}.
Figure (\ref{scalPmeansc}) evidences that the dymanical scaling is well verified for $r \lesssim \tilde{\ell}(t)$,  while at larger distances $\tilde{P}(r,t)$ is characterized by a superposition of slowly decaying step functions as predicted by (\ref{msqav1}).
Figure (\ref{r2meansc}) evidences by means of Montecarlo simulations that 
Equations (\ref{msqav}) are well verified for large times in the whole range 
of $\alpha$'s.

\section*{Acknowledgements}
We acknowledge useful discussion with
P. Barthelemy, J. Bertolotti, R. Livi, D.S. Wiersma, K. Vynck.
This work is partially supported by the MIUR project 
PRIN 2008 \textit{Non linearity and disorder in classical and quantum processes}.

\end{document}